\documentstyle[12pt,epsfig]{article}
     \def\lsim{\raise0.3ex\hbox{$<$\kern-0.75em\raise-1.1ex\hbox{$\sim$}}}
\def\gsim{\raise0.3ex\hbox{$>$\kern-0.75em\raise-1.1ex\hbox{$\sim$}}}
\def\noi{\noindent}
\def\nn{\nonumber}
\def\bea{\begin{eqnarray}}  \def\eea{\end{eqnarray}}
\def\beq{\begin{equation}}   \def\eeq{\end{equation}}

\def\beeq{\begin{eqnarray}} \def\eeeq{\end{eqnarray}}
\def\bmini{\setcounter{hran}{\value{equation}}
     \refstepcounter{hran}\setcounter{equation}{0}
     \renewcommand{\theequation}{\thehran\alph{equation}}\begin{eqnarray}}

\def\bminiG#1{\setcounter{hran}{\value{equation}}
\refstepcounter{hran}\setcounter{equation}{-1}
\renewcommand{\theequation}{\thehran\alph{equation}}
\refstepcounter{equation}\label{#1}\begin{eqnarray}}

\def\emini{\end{eqnarray}\relax\setcounter{equation}{\value{hran}}\ren
ewcommand{\theequation}{\thesection.\arabic{equation}}}

\def\ben{\begin{enumerate}}  \def\een{\end{enumerate}}
    \def\cite#1{[\ref{#1}]}
    
    \def\citt#1#2#3{[\ref{#1},\ref{#2},\ref{#3}]}
    \def\citm#1#2{[\ref{#1}--\ref{#2}]}

\begin{document}
\begin{center}
{\Large \bf Strange particle production at RHIC in the Dual Parton Model} \\

\vskip 8 truemm
{\bf A. Capella$^{\rm a)}$, C.A. Salgado$^{\rm b)}$, D. Sousa$^{\rm c)}$}\\
\vskip 5 truemm

$^{\rm a)}$ Laboratoire de Physique Th\'eorique\footnote{Unit\'e Mixte de
Recherche UMR n$^{\circ}$ 8627 - CNRS}
\\ Universit\'e de Paris XI, B\^atiment 210,
F-91405 Orsay Cedex, France \\

\vskip 3 truemm

$^{\rm b)}$ Theory Division, CERN, CH-1211 Geneva 23, Switzerland \\

\vskip 3 truemm
$^{\rm c)}$ ECT*, Trento, Italy
\end{center}

\begin{abstract}
We compute the mid-rapidity densities of pions, kaons, baryons and 
antibaryons in $Au$-$Au$ collisions at
$\sqrt{s} =$ 130 GeV in the Dual Parton Model supplemented with final 
state interaction (comovers
interaction).  The ratios $B/n_{part}$ ($\overline{B}/n_{part}$) 
increase between peripheral ($n_{part} = 18$) and central ($n_{part} =
350)$ collisions by a factor 2.4 (2.0) for $\Lambda$'s, 4.8 (4.1) for 
$\Xi$'s and 16.5 (13.5) for $\Omega$'s. The ratio $K^-/\pi^-$
increases by a factor 1.3 in the same centrality range. A comparison 
with available data is presented.
\end{abstract}

\vskip 1 truecm

\noi LPT Orsay 02-09 \par
\noi CERN-TH/2002-090 \par
\noi ECT* 02-10 \par
\noi February 2002\par
\newpage
\pagestyle{plain}
The enhancement of the ratio of yields of strange baryons and 
antibaryons per participant nucleon, observed
at CERN-SPS, \cite{1r} \cite{2r}, is one of the main results of the 
Heavy Ion CERN program. A description of
these data has been given in \cite{3r} in  the framework of the Dual 
Parton Model
DPM \cite{4r}, supplemented with final state interaction.  The net
baryon yield is computed in the same framework taking into account the
mechanism of baryon stopping, associated with baryon junction transfer
in rapidity \citm{5r}{9rnew}. We use its implementation in \cite{3r}, 
which describes the SPS data. \par

In the absence of nuclear shadowing, the rapidity density of a given 
type of hadron $h$ produced in $AA$
collisions  at fixed impact parameter, is given by \cite{4r}
\cite{9r}

\bea
\label{1e}
{dN^{AA\to h} \over dy}(y,b) &=& n_A(b) \left [ 
N_{h,\mu(b)}^{qq^{P}-q_{v}^{T}} (y) + N_{h,\mu(b)}^{q_{v}^P - 
qq^T} (y) + (2k-2) \
N_{h,\mu(b)}^{q_s - \overline{q}_s}(y)\right ] \nn \\
&&+  \left ( n(b) - n_A(b) \right )  2k \ 
N_{h,\mu(b)}^{q_s-\overline{q}_s} (y) \ . \eea

\noi Here

\beq
\label{2e}
n(b) = \sigma_{pp} A^2 \int d^2s\ T_A(s) \ T_B(b-s)/\sigma_{AA}(b)
\eeq

\noi is the average number of binary collisions and

\beq
\label{3e}
n_A(b) = A \int d^2s \ T_A(s) \left [ 1 - exp (- \sigma_{pp} \ A \ 
T_A(b-s) \right ]/ \sigma_{AA}(b) \ ,
\eeq

\noi is the average number of participant pairs  at fixed impact 
parameter $b$. $P$ and $T$ denote projectile and target nuclei.
$k$ is the average number of inelastic collisions in $pp$ and $\mu(b) 
= k \nu (b)$ with $\nu (b) = n(b)
/n_A(b)$ is the average total number of collisions suffered by each 
nucleon. At $\sqrt{s} = 130$~GeV we have $k
= 2$ \cite{9r}. \par

The $N_{h,\mu (b)}(y)$ in eq. (\ref{1e}) are the rapidity 
distributions of hadron $h$ in each individual
string. In DPM they are given by convolutions of momentum 
distribution and fragmentation
functions\footnote{For pions, we use the same fragmentation functions 
given in \cite{9r}. For simplicity, the
same form is used for kaons. For $p\overline{p}$ pair production we 
take \cite{10r} $x D_{qq}^p(x) = x
D_{qq}^{\overline{p}}(x) \sim (1 - x)^5$ and $x D_q^p(x) = x 
D_q^{\overline{p}}(x) \sim (1 - x)^3$. For the other
baryon species an extra $\alpha_{\rho}(0) - \alpha_{\phi}(0) = 1/2$ 
is added to the power of $(1 - x)$ for each strange quark in the 
baryon \cite{10r}.}.
The first term in (\ref{1e}) is the rapidity distribution in one $NN$ 
collision of $AA$, resulting from the superposition of $2k$ strings, 
multiplied by the
average number of participant pairs. Since in DPM there are two 
strings per inelastic collision, the second term, consisting of 
strings
stretched between sea quarks and antiquarks, makes up for the total 
average number of strings $2k$ $n(b)$.

It was shown in \cite{9r} that eq. (\ref{1e}), supplemented with 
shadowing corrections, leads to values of
charged multiplicities at mid-rapidities as a function of centrality 
in agreement with data, both
at SPS and RHIC. Here we use the same shadowing corrections as in 
ref. \cite{9r} -- leading to the lower edge of the shaded area in
Fig.~4 of \cite{9r}. \\

\noi \underbar{\bf Net Baryon Production}. Let us now consider the 
net baryon production $\Delta B =
B-\overline{B}$. In the standard version of DPM \cite{4r} (or QGSM 
\cite{11r}) the leading baryon results from
the fragmentation of a valence diquark. This component will be called 
diquark preserving (DP). The
stopping observed in $Pb$ $Pb$ collisions at SPS has led to the 
introduction of a new mechanism
based on the transfer in rapidity of the baryon junction 
\citm{5r}{9rnew}. Here we follow the formalism
in \cite{3r} which describes the SPS data. In an $AA$ collision, this 
component, called diquark-breaking (DB), gives the following
rapidity distribution of the two net baryons in a single $NN$ 
collision of $AA$ \cite{3r}

\beq
\label{4e}
\left ( {dN_{DB}^{\Delta B} \over dy} (y) \right )_{\nu(b)} = 
C_{\nu(b)} \left [ Z_+^{1/2} (1 - Z_+)^{\nu(b) - 3/2}
+ Z_-^{1/2} (1 - Z_-)^{\nu(b) - 3/2} \right ] \eeq

\noi where $Z_{\pm} = \exp (\pm y - y_{max})$ and $\nu(b) = n(b)/n_A(b)$. 
$C_{\nu(b)}$ is determined from the
normalization to two at each $b$.  \par

The net baryon rapidity distribution in $AA$ collisions is then given by

\beq
\label{5e}
{dN^{AA \to \Delta B} \over dy}(y,b) = n_A(b) \left [ {1 \over \nu(b)} \left (
{dN^{\Delta B}_{DP} \over dy} (y) \right )_{\nu(b)} + {\nu (b) - 1 \over
\nu (b)} \left ( {dN_{DB}^{\Delta B} \over dy} (y) \right )_{\nu(b)} 
\right ] \ .
\eeq

The physical content of eq. (\ref{5e}) is as follows. Each nucleon 
interacts in average with
$\nu(b)$ nucleons of the other nucleus. It has been argued in 
\cite{3r} that in only one of these collisions
the string junction, carrying the baryon number, follows a valence 
diquark, which fragments according to the DP
mechanism. In the $\nu (b) - 1$ others, the string junction is freed 
from the valence diquark and net baryon production takes place
according to the DB mechanism, eq. (\ref{4e}). In order to conserve 
baryon number, we have to divide by $\nu
(b)$ and multiply by the number of participating nucleons. We obtain 
in this way eq. (\ref{5e})\footnote{In
the numerical calculations we neglect the first term of (\ref{5e}) 
since the $DP$ component gives a very
small contribution at $y^* \sim 0$ and RHIC energies -- about 5\%
of the $DB$ one
for the most central bin where its effect is maximal.}. This equation 
gives the total net baryon density\footnote{In order to conserve
strangeness locally, we have added an extra $1/2 K^+$ and $1/2K^0$ to 
each produced net $\Lambda$ (plus $\Sigma$'s), an extra $K^+$
and $K^0$ to each net $\Xi$ and an extra $3/2 K^+$ and $3/2 K^0$ to 
each net $\Omega$.}. \par

In order to get the relative densities of each baryon and antibaryon 
species we use simple quark counting rules \cite{3r}. Denoting the
strangeness suppression factor by $S/L$ (with $2L+ S = 1$), baryons produced 
out of three sea quarks (which is the case for pair production) are
given the relative weights $$I_3 = 4L^3 : 4L^3 : 12L^2S : 3LS^2 : 
3LS^2 : S^3 \ .$$

\noi  for $p$, $n$,
$\Lambda + \Sigma$, $\Xi^0$, $\Xi^-$ and $\Omega$, respectively. The 
various coefficients of $I_3$ are obtained from the power expansion 
of $(2L +
S)^3$. In order to take into account the decay of $\Sigma^*(1385)$ 
into $\Lambda \pi$, we redefine the relative rate of $\Lambda$'s and
$\Sigma$'s using the empirical rule $\Lambda = 0.6(\Sigma^+ + 
\Sigma^-$) -- keeping, of course, the total
yield of $\Lambda$'s plus $\Sigma$'s unchanged. In this way the 
normalization constants of all baryon species in
pair production are determined from one of them. This constant, 
together with the relative normalization of
$K$ and $\pi$, are determined from the data for very peripheral 
collisions. In the calculations we use $S =
0.1$ $(S/L = 0.22)$. \par

For net baryon production two possibilities have been considered. The 
first one is that the behaviour in $Z^{1/2}$, eq. (\ref{4e}), is
associated to the transfer of the string junction without valence 
quarks \citt{5r}{7r}{9rnew}. In this case the net baryon is made out 
of three sea quarks
and the relevant weights are given by $I_3$. In the second one, eq. 
(\ref{4e}) is a pre-asymptotic term associated to the transfer of
the baryon junction plus one valence quark \cite{6r}. In this 
case the relevant weights are given by $I_2$, i.e. from the various
terms in the expansion of $(2L + S)^2$. This second possibility is 
favored by the data\footnote{Note, however, that a non-zero value
of net omegas has been observed in $hA$ collisions \cite{13rnew}. 
This requires a non-vanishing contribution proportional to $I_3$.
However, its effect in $AA$ collisions is presumably small since, in 
this case, the net omegas are almost entirely due to final state
interactions.}. Since the normalization of the total net baryon yield 
is determined from baryon number
conservation, there is  no extra free normalization constant. 
Moreover, the total net baryon yield is not affected by final state
interaction.\\

\noi \underbar{\bf Final State Interactions}. The hadronic densities 
obtained above will be used as initial
conditions in the gain and loss differential equations which govern 
final state interactions. In the
conventional derivation \cite{12r} of these equations, one uses 
cylindrical space-time variables and assumes
boost invariance. Furthermore, one assumes that the dilution in time 
of the densities is only due to
longitudinal motion, which leads to a $\tau^{-1}$ dependence on the 
longitudinal proper time $\tau$. These
equations can be written as \cite{12r} \cite{3r}

\beq
\label{6e}
\tau {d\rho_i \over d \tau} = \sum_{k\ell} \sigma_{k\ell} \ \rho_k \ 
\rho_{\ell} - \sum_k \sigma_{ik} \ \rho_i
\ \rho_k \ . \eeq

\noi The first term in the r.h.s. of (\ref{6e}) describes the 
production (gain) of particles of type $i$
resulting from the interaction of particles $k$ and $\ell$. The 
second term describes the loss of particles
of type $i$ due to its interaction with particles of type $k$. In eq. 
(\ref{6e}) $\rho_i = dN_i/dy d^2s(y,b)$ are the particles
yields per unit rapidity and per unit of transverse area, at fixed 
impact parameter. They can be obtained from the rapidity densities
(\ref{1e}), (\ref{5e}) using the geometry, i.e. the $s$-dependence of 
$n_A$ and $n$, eqs. (\ref{2e}), (\ref{3e}). The procedure is
explained in detail in \cite{13r}. $\sigma_{k\ell}$ are the 
corresponding cross-sections averaged over the momentum
distribution of the colliding particles. \par

Equations (\ref{6e}) have to be integrated from initial time $\tau_0$ 
to freeze-out time $\tau_f$. They are
invariant under the change $\tau \to c \tau$ and, thus, the result 
depends only on the ratio $\tau_f/\tau_0$.
We use the inverse proportionality between proper time and densities and put
$\tau_f/\tau_0 = (dN/dyd^2s(y,b))/\rho_f$. Here the numerator is 
given by the DPM particles densities. We take
$\rho_{f} = [3/\pi R_p^2](dN^-/dy)_{y^*\sim 0} = 2$~fm$^{-2}$, which 
corresponds to the charged density per
unit rapidity in a $pp$ collision at $\sqrt{s} = 130$~GeV. This 
density is about 70 \% larger \cite{9r} than
at SPS energies. Since the corresponding increase in the $AA$ density 
is comparable, the average duration
time of the interaction will be approximately the same at CERN SPS 
and RHIC -- about 5 to 7 fm.\par

Next, we specify the channels that have been taken into account in 
our calculations. They are

\beq
\label{7e}
\pi N \stackrel{\rightarrow}{\leftarrow} K \Lambda (\Sigma)\ , \quad 
\pi \Lambda (\Sigma )
\stackrel{\rightarrow}{\leftarrow} K \Xi \ , \quad \pi \Xi 
\stackrel{\rightarrow}{\leftarrow} K \Omega  \eeq

\noi We have also taken into account the strangeness exchange reactions

\beq
\label{8e}
\pi \Lambda (\Sigma ) \stackrel{\rightarrow}{\leftarrow} K N\ , \quad \pi \Xi
\stackrel{\rightarrow}{\leftarrow} K \Lambda (\Sigma ) \ , \quad \pi 
\Omega  \stackrel{\rightarrow}{\leftarrow}
K \Xi \ . \eeq

\noi as well as the channels corresponding to (\ref{7e}) and 
(\ref{8e}) for antiparticles\footnote{To be precise, of all possible 
charge combinations in reactions (\ref{7e}), we have only kept those
involving the annihilation of a light $q$-$\overline{q}$ pair and 
production of an $s$-$\overline{s}$ in the
$s$-channel. The other reactions, involving three quarks in the 
$t$-channel intermediate state, have
substantially smaller cross-sections and have been neglected. All 
channels involving $\pi^0$ have been taken with
cross-section $\sigma /2$ since only one of the $u\overline{u}$ and 
$d\overline{d}$ components
of $\pi^0$ can participate to a given charge combination. For details 
see the second paper of \cite{3r}.}. We have taken $\sigma_{ik} = 
\sigma =
0.2$~mb, i.e. a single value for all reactions in (\ref{7e}) and 
(\ref{8e}) -- the same value
used in ref. \cite{3r} to describe the CERN SPS data. \\

\noi \underbar {\bf Numerical Results}. All our results refer to 
mid-rapidities. The calculations have been
performed in the interval $-0.35 < y^* < 0.35$. In Fig.~1a-1d we show 
the rapidity densities of
$B$, $\overline{B}$ and $B - \overline{B}$\footnote{A Monte Carlo 
calculation in a
similar framework with string fusion can be found in \cite{14rnew}. A 
net proton rapidity density of about 10 for central $Au$ $Au$
collisions at mid-rapidities at RHIC was first predicted in \cite{8r} 
using a stopping mechanism similar to the one considered here.} versus
$h^- = dN^-/d\eta = (1/1.17) dN/dy$ and compare them with available 
data \citm{14r}{16r}. We see that, in first approximation, $p$, 
$\overline{p}$,
$\Lambda$ and $\overline{\Lambda}$ scale with $h^-$. Quantitatively, 
there is a slight decrease with centrality of
$p/h^-$ and $\overline{p}/h^-$ ratios, a slight increase of $\Lambda 
/h^-$ and $\overline{\Lambda}/h^-$ and a much larger increase for 
$\Xi$
($\overline{\Xi})/h^-$ and $\Omega$ ($\overline{\Omega})/h^-$. In 
Fig.~2a and 2b we plot the yields of $B$ and $\overline{B}$ per 
participant normalized
to the same ratio for peripheral collisions versus $n_{part}$. The 
enhancement of $B$ and $\overline{B}$ increases with the number of 
strange quarks in
the baryon. This increase is comparable 
to the one found at SPS between pA and
central PbPb collisions -- somewhat larger for antibaryons.
The ratio $K^-/\pi^-$ increases by 30~\% in the same 
centrality range, between 1.1 and 1.4 in agreement with present data
\cite{14r}. The ratios $\overline{B}/B$ have a mild decrease with 
centrality of about
15~\% for all baryon species -- which is also seen in the data 
\cite{18r}. Our values for $N^{ch}/N_{max}^{ch} = 1/2$ are~:
$\overline{p}/p = 0.69$, $\overline{\Lambda}/\Lambda = 0.72$, 
$\overline{\Xi}/\Xi = 0.79$, $\Omega/\overline{\Omega} = 
0.83$\footnote{In ref. \cite{3r},
the relative weights of net baryons were given by the factors 
$0.5(I_2 + I_3)$ -- instead of $I_2$. In this case the values of the 
ratios are
$\overline{p}/p = 0.70$, $\overline{\Lambda}/\Lambda = 0.71$, 
$\overline{\Xi}/\Xi = 0.76 $ and $\overline{\Omega}/\Omega = 0.78$.
Their increase with the number of
strange quarks in the baryon
is smaller.}, to be compared with the 
measured values \cite{15r}~: $$\overline{p}/p = 0.63 \pm 0.02 \pm 
0.06 \quad , \quad
\overline{\Lambda}/\Lambda = 0.73 \pm 0.03 \quad , \quad 
\overline{\Xi}/\Xi = 0.83 \pm 0.03 \pm 0.05 \ .$$

\noi The ratio $K^+/K^- = 1.1$ and has a mild increase with 
centrality, a feature also seen in the data. \par

As explained above only one parameter has been adjusted in order to 
determine the absolute yields of baryons and antibaryons. Since both 
the PHENIX
\cite{14r} and the STAR data \cite{16r} are not corrected for 
feed-down from weak decays, this free parameter has to be 
re-determined after these
corrections are known. It will then be possible to compare its value 
with the one obtained from other sets of data, in particular $pp$.
Likewise, the corrected yields of net protons will determine the 
exact amount of stopping and should allow to decide whether or not there is 
``anomalous''
stopping in $AA$, i.e. an excess as compared to an extrapolation from 
$pp$ and $pA$. At present there is no clear indication of anomalous 
stopping.\\

After completion of this work, 
the PHENIX collaboration \cite{21r} has published the yields of $p$ 
and $\overline{p}$
for the 5~\% most central events corrected for feed-down.The corrections
are of 30~\%.The systematic error is 20~\%.Also, C.Roy
(STAR collaboration)
communicated to us a recent STAR preliminary result on the
mid-rapidity
densities of $\Xi^-$ and $\Xi^+$ for the 14~\% most central events.
These values are 60 to 70~\% higher than our results. Their
systematic error is
20~\%.\par

\noi \underbar{\bf Physical interpretation}. Before final state 
interactions, all ratios $K/h^-$, $B/h^-$
and $\overline{B}/h^-$ decrease slightly with increasing centrality. 
This effect is rather marginal at RHIC
energies and mid-rapidities. \par

The final state interactions (\ref{7e}), (\ref{8e}) lead to a gain of 
strange particle
yields.  The reason for this is the following. In the first direct 
reaction (\ref{7e}) we have $\rho_{\pi} >
\rho_K$, $\rho_N > \rho_{\Lambda}$, $\rho_{\pi} \rho_N \gg \rho_K 
\rho_{\Lambda}$. The same is true for all
direct reaction (\ref{7e}). In view of that, the effect of the 
inverse reactions (\ref{7e}) is small. On the contrary, in
all reactions (\ref{8e}), the product of densities in the initial and 
final state are comparable and the
direct and inverse reactions tend to compensate with each other. 
Baryons with the largest strange quark
content, which find themselves at the end of the chain of direct 
reactions (\ref{7e}) and have the smallest
yield before final state interaction, have the largest enhancement. 
Moreover, the gain in the yield of
strange baryons is larger than the one of antibaryons since $\rho_B > 
\rho_{\overline{B}}$. Furthermore, the
enhancement of all baryon species increases with centrality, since 
the gain, resulting from the first term in
eq. (\ref{6e}), contains a product of densities and thus, increases 
quadratically with increasing centrality.\par

Although the inverse slopes (``temperature'') have not been discussed 
here, let us note that in DPM they are approximately the same for
all baryons and antibaryons both before and after final 
state interaction -- the effect of final state interaction being
rather small \cite{14rnew} \cite{20r}. \par

In conclusion, we would like to emphasize the fact that in DPM 
(before final state interaction) the
rapidity density of charged particle per participant increases with 
centrality. 
This increase is
larger for low centralities \cite{9r}.
This has an important effect on both the size and the
pattern of strangeness enhancement in our results. 
It explains why the
departure from a linear increase of $\Xi$'s and $\Omega$'s (concave 
shape) seen in Figs.~1c and 1d is also more pronounced for lower
centralities. It leads to the convexity in the centrality 
dependence of the yields of hyperons and antihyperons per
participant in Figs. 2 (Note, however, a
change of curvature for very peripheral collisions where the effect of
final state interaction is negligible). This centrality pattern is a distinctive 
feature as well as a firm prediction
of our approach.\\

\noi {\large\bf Acknowledgments}\par
It is a pleasure to thank N. Armesto, A. Kaidalov, 
K. Redlich and Yu. Shabelski 
for discussions and C. Roy (STAR) and M.J. Tannenbaum (PHENIX) for 
information on the
data. C.A.S. is supported by a Marie Curie
Fellowship of the European Community program TMR (Training and Mobility of
Researchers), under the contract number HPMF-CT-2000-01025.\\

\newpage

\noindent
{\large\bf References}

\ben
\item\label{1r} WA97 coll., E. A. Andersen et al, Phys. Lett. {\bf 
B433}, 209 (1998)~; Phys. Lett. {\bf B449}, 401 (1999). \\
NA57 coll., N. Carrer, Nucl. Phys. {\bf A698}, 118c (2002).
 
\item\label{2r} NA49 coll., H. Appelsh\"auser et al, Phys. Lett. {\bf 
B433}, 523 (1998)~; Phys. Lett. {\bf B444}, 523 (1998)~; Eur. Phys.
J. {\bf C2}, 661 (1998).
 
\item\label{3r} A. Capella and C. A. Salgado, New Journal of Physics 
{\bf 2}, 30.1-30.4 (2000)~; Phys. Rev. {\bf C60}, 054906 (1999). \\
A. Capella, E. G. Ferreiro and C. A. Salgado, Phys. Lett. {\bf B459}, 
27 (1999)~; Nucl. Phys. {\bf A661}, 502 (1999).

\item\label{4r} A. Capella, U. Sukhatme, C.-I. Tan and J. Tran Thanh 
Van, Phys. Lett. {\bf 81B}, 68 (1979)~; Phys. Rep. {\bf 236}, 225
(1994).

\item\label{5r} G. C. Rossi and G. Veneziano, Nucl. Phys. {\bf B123}, 
507 (1977).
 
\item\label{6r} B. Z. Kopeliovich and B. G. Zakharov, Sov. J. Nucl. 
Phys. {\bf 48}, 136 (1988)~; Z. Phys. {\bf C43}, 241 (1989).
 
\item\label{7r} D. Kharzeev, Phys. Lett. {\bf B378}, 238 (1996).

\item\label{8r} A. Capella and B. Z. Kopeliovich, Phys. Lett. {\bf B381}, 
325 (1996).

\item\label{9rnew} S. E. Vance and M. Gyulassy, Phys. Rev. Lett. {\bf 
83}, 1735 (1999).
 
\item\label{9r} A. Capella and D. Sousa, Phys. Lett. {\bf B511}, 185 (2001).
 
\item\label{10r} G. H. Arakelyan, A. Capella, A. Kaidalov and Yu. M. 
Shabelski, hep-ph/0103337, to be published in Z. Phys. C.
 
\item\label{11r} A. B. Kaidalov, Phys. Lett. {\bf 116B}, 459 (1982).

\item\label{13rnew} E769 coll., E. M. Aitala et al., 
Phys. Lett. {\bf B496}, 9 (2000).

\item\label{12r} B. Koch, U. Heinz and J. Pitsut, Phys. Lett. {\bf 
B243}, 149 (1990).

\item\label{13r} A. Capella, A. B. Kaidalov and D. Sousa, 
nucl-th/0105021, to be published in Phys. Rev. C.

\item\label{14rnew} N. S. Amelin, N. Armesto, C. Pajares and D. 
Sousa, Eur. Phys. J. C {\bf 22}, 149 (2001).

\item\label{14r} PHENIX coll., K. Adcox et al, nucl-ex/0112006.

\item\label{15r} STAR coll., C. Roy, nucl-ex/111017.

\item\label{16r} STAR coll., C. Adler et al, Phys. Rev. Lett.
{\bf 87}, 262302 (2001).

\item\label{17r} STAR coll., H. Caines, Nucl. Phys. {\bf A698}, 112c (2002).

\item\label{18r} STAR coll., C. Adler et al, Phys. Rev. Lett. {\bf 86}, 4778 (2001).

\item\label{20newr} STAR coll., J. Castillo, Proc. Strangeness in 
Quark Mater 2001, Frankfurt (Germany) to be published.

\item\label{20r} J. Ranft, A. Capella and J. Tran Thanh Van, Phys. 
Lett. {\bf B320}, 346 (1994).

\item\label{21r} PHENIX coll., K.Adcox et al, nucl-ex/0204007.

\een

\vskip 2 truecm
\centerline{\large \bf Figure Caption :}
\vspace{0.3cm}

\noi {\bf Figure 1.}  (a) Calculated values of rapidity densities of 
$p$ (solid line), $\overline{p}$ (dashed line), and $p -
\overline{p}$ (dotted line) at mid rapidities, $|y^*| < 0.35$, are 
plotted as a function of $dN^-/d\eta$, and
compared with PHENIX data \cite{14r}~; (b) same for $\Lambda$ and 
$\overline{\Lambda}$ compared to preliminary STAR data \cite{16r}~; 
(c)
same for $\Xi^-$ and $\overline{\Xi}^+$~; (d) same for $\Omega$ and 
$\overline{\Omega}$.\\

\noi {\bf Figure 2.} Calculated values of the ratios $B/n_{part}$ (a) 
and $\overline{B}/n_{part}$ (b), normalized
to the same ratio for peripheral collisions ($n_{part} = 18$), are plotted as a 
function of $n_{part}$.

\newpage
\centerline{\bf Fig. 1}
\begin{figure}[hbtp]
\begin{center}
\mbox{\epsfig{file=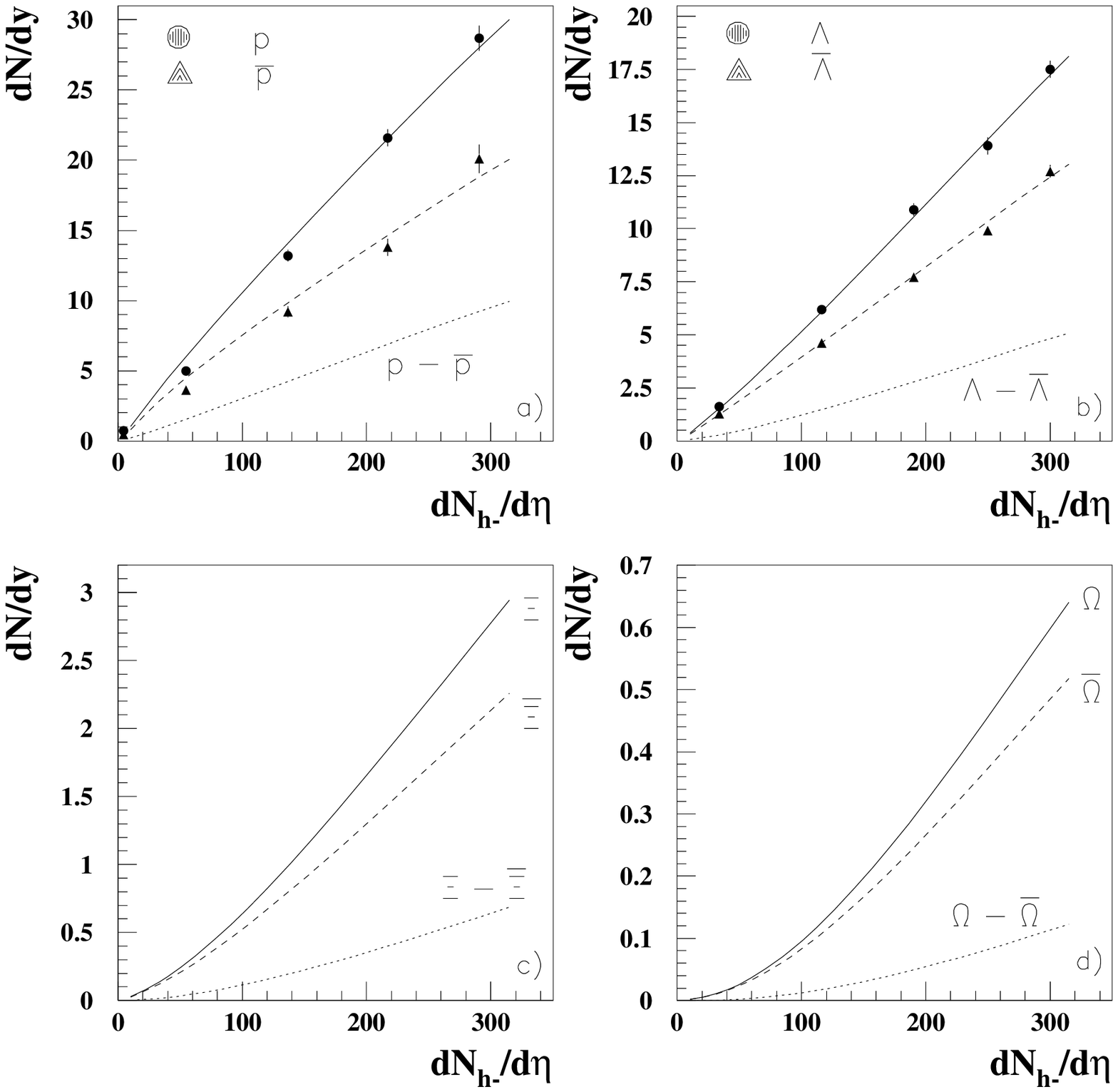,height=16cm}}
\end{center}
\end{figure}

\newpage
\centerline{\bf Fig. 2.a}
\begin{figure}[hbtp]
\begin{center}
\mbox{\epsfig{file=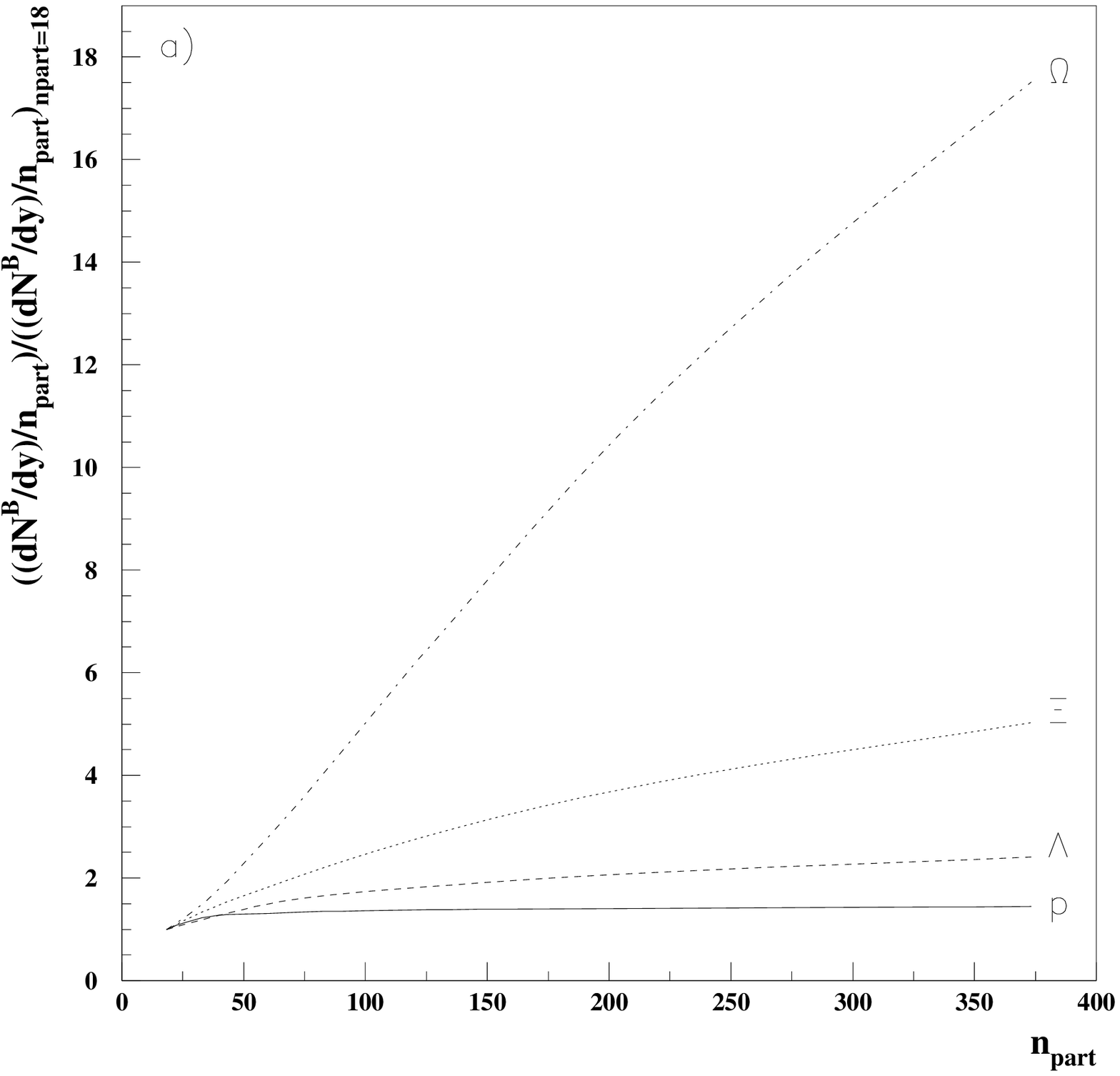,height=16cm}}
\end{center}
\end{figure}

\newpage
\centerline{\bf Fig. 2.b}
\begin{figure}[hbtp]
\begin{center}
\mbox{\epsfig{file=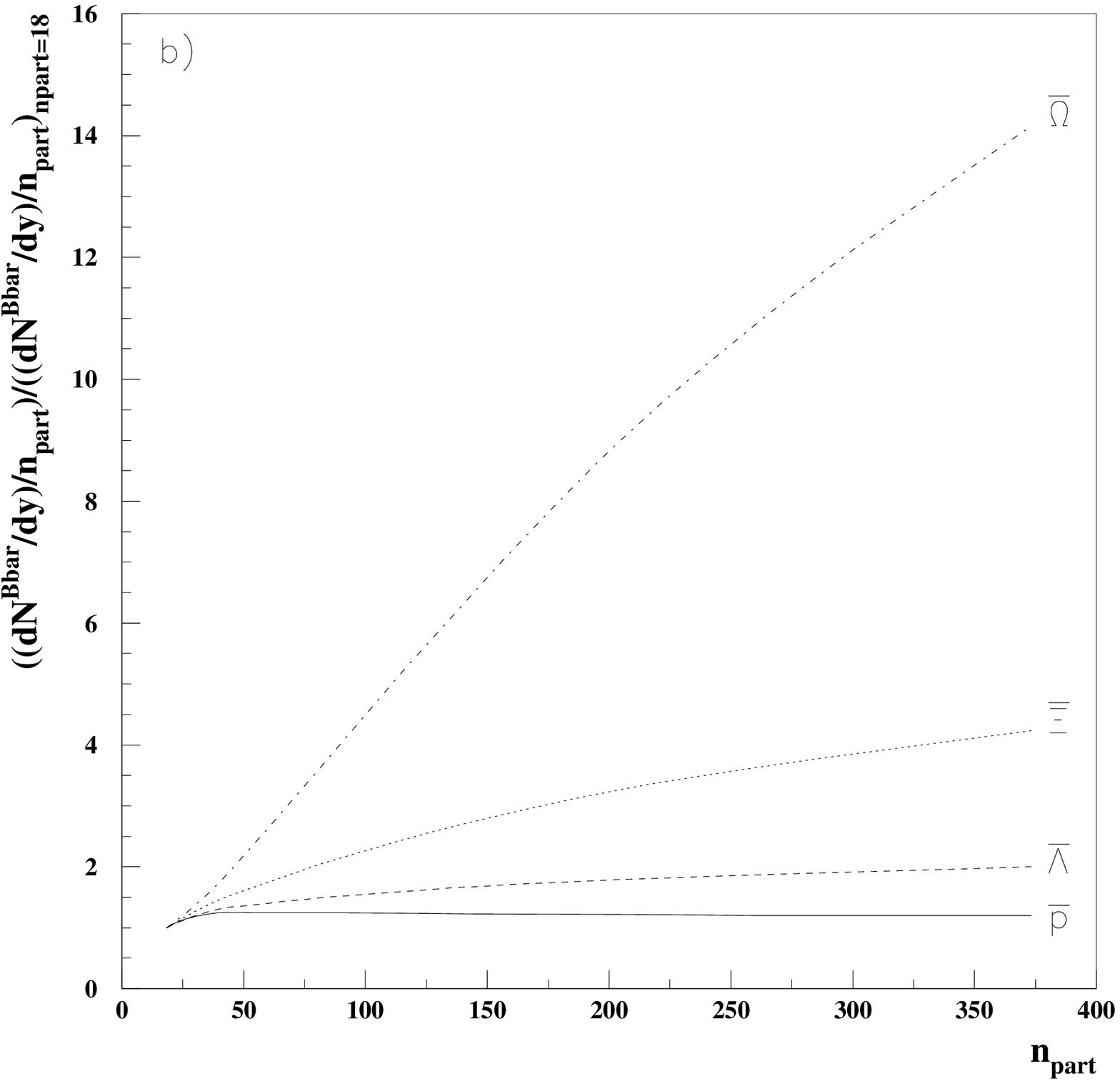,height=16cm}}
\end{center}
\end{figure}

\end{document}